\documentclass[a4paper]{article}
\usepackage{INTERSPEECH2020}
\usepackage{multirow}
\usepackage{adjustbox}
\usepackage{amssymb}
\usepackage{mathtools}
\usepackage{arydshln}
\usepackage{url}
\usepackage{cite}

\title{Acoustic Scene Classification using Audio Tagging}
\name{Jee-weon Jung, Hye-jin Shim, Ju-ho Kim, Seung-bin Kim, and Ha-Jin Yu$^\dag$\thanks{$^\dag$ Corresponding author}}
\address{School of Computer Science, University of Seoul, Republic of Korea}
\email{jeewon.leo.jung@gmail.com, shimhz6.6@gmail.com, wngh1187@naver.com, \\kimho1wq@naver.com, hjyu@uos.ac.kr}

\begin{document}

\maketitle
\begin{abstract}
Acoustic scene classification systems using deep neural networks classify given recordings into pre-defined classes.  
In this study, we propose a novel scheme for acoustic scene classification which adopts an audio tagging system inspired by the human perception mechanism.  
When humans identify an acoustic scene, the existence of different sound events provides discriminative information which affects the judgement.  
The proposed framework mimics this mechanism using various approaches.  
Firstly, we employ three methods to concatenate tag vectors extracted using an audio tagging system with an intermediate hidden layer of an acoustic scene classification system.  
We also explore the multi-head attention on the feature map of an acoustic scene classification system using tag vectors.  
Experiments conducted on the detection and classification of acoustic scenes and events 2019 task 1-a dataset demonstrate the effectiveness of the proposed scheme.  
Concatenation and multi-head attention show a classification accuracy of $75.66$ \% and $75.58$ \%, respectively, compared to $73.63$ \% accuracy of the baseline.  
The system with the proposed two approaches combined demonstrates an accuracy of $76.75$ \%.  

\end{abstract}
\noindent\textbf{Index Terms}: acoustic scene classification, audio tagging, attention

\section{Introduction}
\label{sec:intro}
Acoustic scene classification (ASC) is an emerging task with a wide range of applications. 
The task is to identify a given audio recording into one of the pre-defined acoustic scenes, i.e., classes. 
Leveraging the recent advances in deep learning, majority of ASC systems utilize deep neural networks (DNNs) \cite{Koutini2019, Huang2019, Hyeji2019}. 
To facilitate the studies on the ASC task, the detection and classification of acoustic scenes and events (DCASE) community is providing a platform with annual competitions and public datasets \cite{DCASE2017Workshop, DCASE2018Workshop, DCASE2019Workshop}. 

Most ASC systems perform the task either in an end-to-end manner of classifying the input recording directly or by using a DNN’s last hidden layer output as a representation vector and utilize a back-end classifier \cite{Zheng2019, Jung2018DNN}. 
Both methodologies utilize a single step approach. 
On the other hand, humans are known to first recognize the existence of different sound events and then conduct scene classification \cite{guastavino2007categorization}. 
This implies that humans conduct ASC task using a two-step approach. 
For example, the existence of an airplane takeoff sound event can be used to infer that the scene is an airport and not a shopping mall.

\begin{figure}[t]
  \centering
  \includegraphics[width=\linewidth]{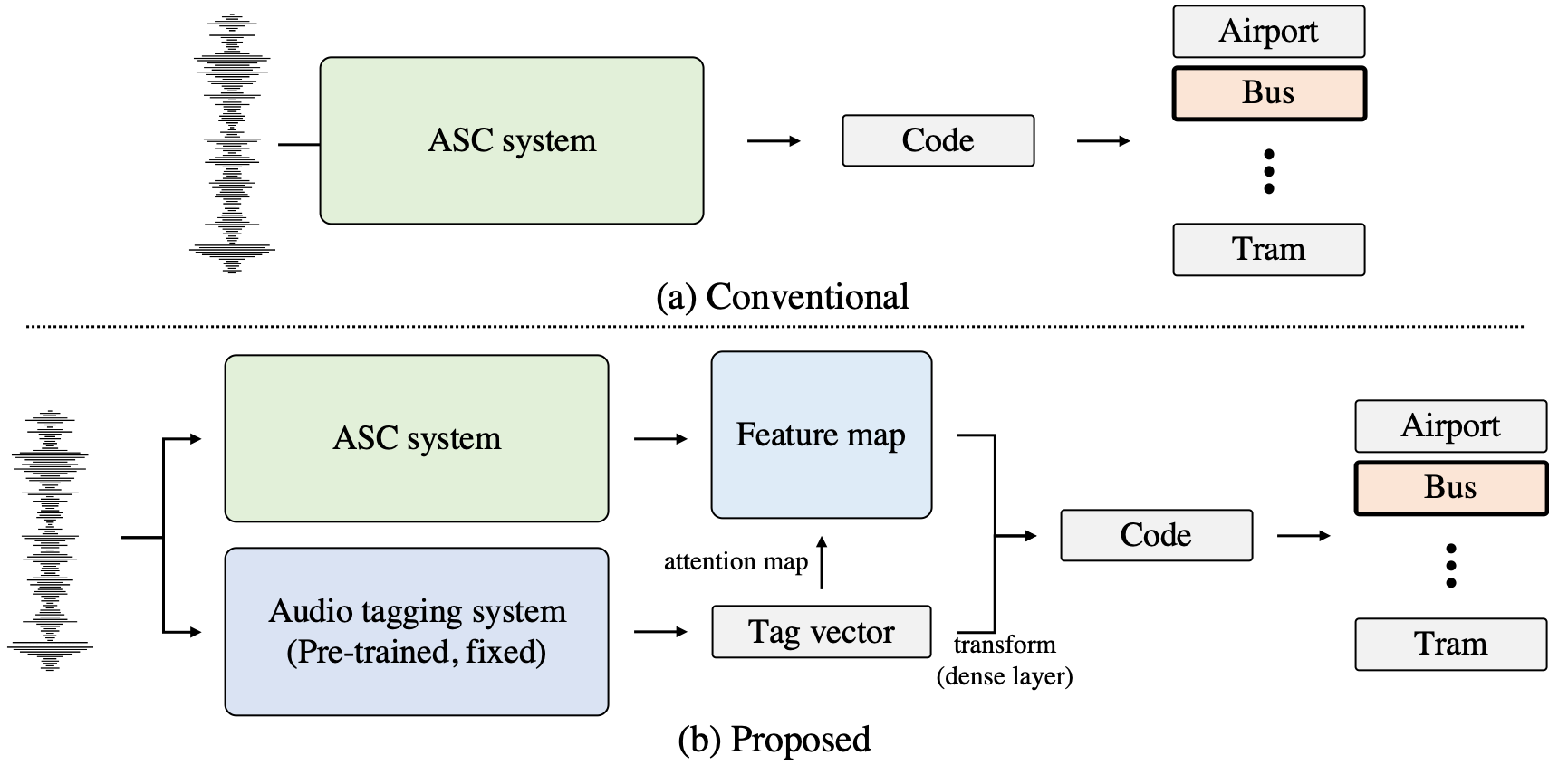}
  \caption{Conceptual illustration of \textbf{(a)}: conventional ASC system and \textbf{(b)}: the proposed scheme using audio tagging.}
  \label{fig:concept}
\end{figure}

Recently there have been few studies to jointly conduct the ASC and the audio tagging task \cite{bear2019towards, imoto2020sound, imoto2016acoustic}.  
Bear \textit{et. al.} \cite{bear2019towards} proposed to jointly conduct sound event detection and ASC task where ASC is conducted by a majority voting mechanism based on the result of sound event detection.  
Imoto \textit{et. al.} \cite{imoto2016acoustic} studied the improvement of sound event detection performance by using a multi-task learning of sound events and acoustic scenes. 
It investigated the use of sound event information for the ASC task by proposing a graphical model that sequentially conducts event detection and scene classification.  
To our knowledge, however, this paper is the first to use tag vectors extracted from a separate audio tagging system to improve the ASC system.

In this paper, we propose an ASC system inspired by the human perception mechanism and utilizes tag vectors.  
We expect that the ASC system can improve by using the information from the tag vectors.  
A tag vector refers to the output of an audio tagging system that represents the existence of various sound events.  
Figure \ref{fig:concept} illustrates the concept of the proposed scheme compared to conventional DNN-based ASC systems.  
The proposed framework utilizes a trained audio tagging system that demonstrated the best performance in DCASE 2019 task \cite{Akiyama2019}.  
A given recording is input to both the audio tagging and the ASC system to extract a tag vector and a representation vector, i.e. code.  
Both the tag vector and the code are used for the ASC task. 
We propose various methods to utilize the tag vectors for the ASC task.

Specifically, we first propose to concatenate a tag vector with a code of the ASC task.  
Three methods are addressed in Section \ref{sec:proposed}.1, and are explored to derive a concatenated representation and experimentally validated.  
We also propose to apply the attention mechanism to the feature map of the ASC task using a tag vector.  
An attention mechanism exclusively emphasizes a DNN's intermediate representation with or without exploiting external representations \cite{bahdanau2014neural, vaswani2017attention, chan2016listen,safari2019self, Ren2018}.  
Our proposal belongs to the former case where a feature map is attended by an attention map derived from a tag vector.  
We also explore multi-head attention \cite{vaswani2017attention} mechanism which demonstrates competitive performances.  
Finally, we achieve further improvements by combining the two proposed approaches.  

The rest of this paper is organized as follows. 
Section \ref{sec:baselines} describes the two baseline systems, an ASC system and an audio tagging system. 
Section \ref{sec:proposed} introduces the proposed framework which utilizes tag vectors for the ASC task. 
Experimental validation of the proposed approach is presented in Section \ref{sec:exp} and the paper is concluded. 

\begin{figure*}[t]
  \centering
  \includegraphics[width=\linewidth]{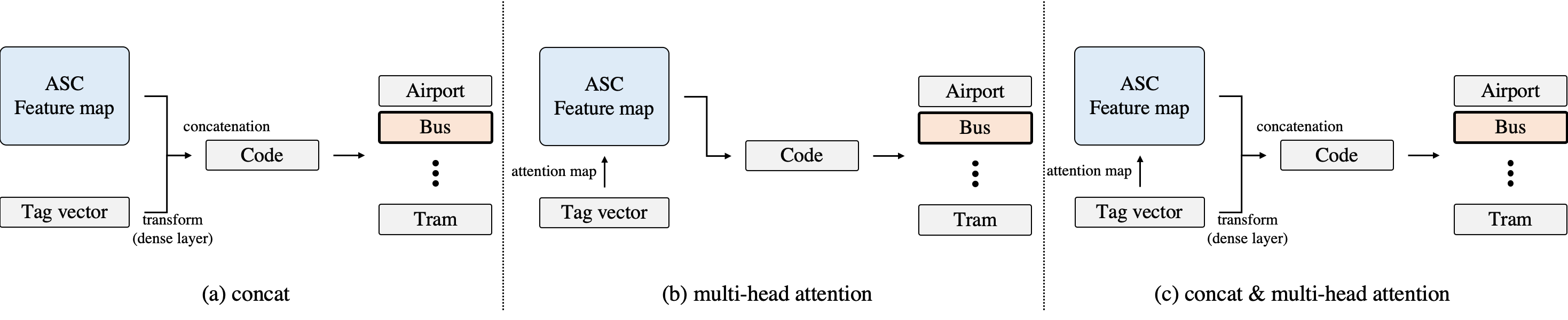}
  \caption{Proposed methods which utilize tag vectors for the ASC task. \textbf{(a)}: Concatenates the tag vector with the output of ASC feature map. Concatenating with or without feature transformation is explored. \textbf{(b)}: Conduct a multi-head attention where the tag vector is used to derive an attention map for the ASC feature map. \textbf{(c)}: Both (a) and (b) are conducted.} 
  \label{fig:overall}
\end{figure*}

\section{Baseline systems}
\label{sec:baselines}

\begin{table}[t]
 \caption{DNN architecture of the baseline. Strided-conv refers to a convolutional layer that has a stride size identical to filter length for processing raw waveforms \cite{Jung2017}. Numbers inside Conv refer to filter length, stride, and number of filters. (BN: batch normalization \cite{BatchNormalization}, FC: fully-connected.)} 
  \centering
  \label{tab:dnn_arc}
  \begin{tabular}{l c c}
  \toprule
  \textbf{Layer} & \textbf{Input:(479999, 2)} & \textbf{Output shape}\\
  \toprule
  \multirow{2}{*}{Strided} & Conv(12,12,128) & \multirow{3}{*}{(39999, 128)}\\
  \multirow{2}{*}{-conv}& BN & \\
  & LeakyReLU & \\
  \midrule
  Res block & 
    $\left \{
      \begin{tabular}{c}
      Conv(3,1,128)\\
      BN \\
      LeakyReLU\\
      Conv(3,1,128)\\
      BN\\
      LeakyReLU\\
      \hdashline
      MaxPool(3)\\
      \end{tabular}
    \right \}$
    $\times$7
    
  & (18, 128)\\
  \midrule
  Gl avg & Global avg pool() & (128,)\\
  GL max & Global max pool() & (128,)\\
  Concat & - & (256,)\\
  \midrule
  Code & FC(64) & (64,)\\
  \midrule
  Output & FC(10) & (10,)\\
  \bottomrule
  \end{tabular}
\end{table}

\subsection{ASC system}
The ASC system used for this paper comprises two sub-systems: a front-end DNN for feature (code) extraction, and a back-end support vector machine (SVM) for classification.  
We adopt a front-end DNN that directly inputs raw waveforms \cite{Jung2018DNN, jung2019distilling}.  
The system first extracts frame-level representations using convolutional layers with residual connections \cite{he2016deep, he2016identity}.  
A global max pooling and a global average pooling aggregates frame-level features, and they are fed to a fully-connected layer. The output of the layer is used as the code.  
The acoustic scenes are classified in the train phase, but the output layer is removed after training.  
Table \ref{tab:dnn_arc} describes the overall architecture.  
 
In this study, an SVM with radial basis function or sigmoid kernel is used for back-end classification.  
SVM is widely selected as a back-end classifier of the ASC task.  
Unless otherwise stated, all performance reported in this paper is of the systems using SVM as the back-end classification.

\subsection{Audio tagging system}
The multi-label audio tagging task determines whether various defined sound events exist in an input audio recording.  
In this task, the output is a vector of dimension equal to the number of pre-defined sound events. Each dimension has a value within the range of $0$ to $1$ and describes the posterior probability of a sound event using a sigmoid non-linear function.  
Throughout this paper, the output vector will be referred to as `tag vector'.  
In this study, we use the system proposed by Akiyama \textit{et. al.} \cite{Akiyama2019}, the winning system of the DCASE 2019 challenge task 2, which adopts a multi-task learning architecture and soft pseudo-label framework.   
The system aims to train a multi-label audio tagging system under a scenario where a small portion of human labeled data and abundant auto-generated noisy labels exist.  
Specifically, we use the system with Mel-spectrogram input and ResNet \cite{he2016deep} architecture combination as it is the best performing single system reported. 
Further details regarding this system are provided in \cite{Akiyama2019}.

\section{Audio tagging-based ASC}
\label{sec:proposed}
In this section we propose two approaches that use a tag vector for the ASC task.  
Tag vector in this paper refers to an output of an audio tagging system where each dimension represents the existence of a pre-defined sound event in a real value of $0$ to $1$.  
The two proposed methods can also be used concurrently, leading to further improvements.  
Figure \ref{fig:overall} illustrates two methods and the final proposed system using both methods.  
Figure \ref{fig:overall}-(a)shows that the first method concatenates a tag vector with an intermediate representation of the ASC system.  
Figure \ref{fig:overall}-(b) shows that the second method uses a multi-head attention using tag vectors.  
The use of the two methods concurrently is depicted in Figure \ref{fig:overall}-(c).  

\subsection{Concatenation}
In this sub-section, three methods are addressed to concatenate a tag vector with the feature map of an ASC system (Figure \ref{fig:overall}-(a)).  
The simplest approach is to concatenate the feature map of an ASC system after global pooling (`\textit{Code}' in Table \ref{tab:dnn_arc}) with a tag vector and use it as a code representation.  
However, this approach increases the dimensionality of the code and the two representations separately exist in each subspace of the concatenated representation. 
The second method conducts feature transformation using a fully-connected layer after concatenating the code of ASC and a tag vector.  
In this case, the transformed representation is used as the code.  
 
Additionally, we propose to use a feature transformation to a tag vector before concatenation. 
A tag vector itself represents the probability of each pre-defined sound events.  
We hypothesize that the transformations before concatenation via fully-connected layers would lead to more discriminative features.  
Experiments in Table \ref{tab:concat} show that all three approaches show improvements compared to the baseline where the third approach demonstrated the best performance.

\subsection{Multi-head attention}
Attention is a widely used mechanism that exclusively emphasizes discriminative features \cite{bahdanau2014neural, vaswani2017attention, chan2016listen,safari2019self, Ren2018}. 
It utilizes a vector referred to an attention map where each value of an attention map has a value between $0$ and $1$, summing up to $1$ using a softmax function. 
Attention is conducted by multiplying a given feature with an attention map. 
This mechanism has been actively adopted for the ASC task, where self-attention scheme which does not utilize external data was mainly studied. 
In this study, we propose to use a multi-head attention on the filter dimension of a feature map of the ASC system where a tag vector is used to derive the attention vector.  
Multi-head attention divides the representation space (filters in this case) into the same number of sub-spaces as the number of heads. It then applies softmax function to each sub-space, whereas conventional attention mechanisms apply softmax function to the entire space \cite{vaswani2017attention}.  
The assumption based on the use of attention with tag vectors is that the information regarding sound events that reside in tag vectors is sufficient to emphasize discriminative filters for the ASC task.

Let $M$ be a feature map of an ASC task, $M \in \mathbb{R}^{f\times t}$ described as:
\begin{equation}
    M=\begin{bmatrix}
    M_{11} & M_{12} & \dotsb\\
    \vdots & \ddots & \\
    M_{h1} & \dotsb & M_{ht}\\
\end{bmatrix}, M_{ht} \in \mathbb{R}^{f/h},
\end{equation}
where $f$ refers to the number of filters, $t$ refers to the length of the sequence in time dimension, and $h$ refers to the number of heads. 
Let $T$ be a tag vector, $T \in \mathbb{R}^{c}$ where $c$ refers to the number of pre-defined sounds events in the audio tagging task.  
We first derive an attention map $A \in \mathbb{R}^f$, using a tag vector through feature transformation using fully-connected layers.  
Then we split the attention map into $h$ which refers to the number of heads and then subsequently apply softmax function and then concatenate again.  
Derived attention map is denoted as $A = [A_1, A_2, \dotsb, A_h]$, $A_h \in \mathbb{R}^{f/h}$. 
We calculate the attention-applied feature map, $M'$ as: 
\begin{equation}
    M' = \begin{bmatrix}
    M_{11}\cdot M_1 & M_{12}\cdot M_1 & \dotsb\\
    \vdots & \ddots & \\
    M_{h1}\cdot M_h & \dotsb & M_{ht}\cdot am_h\\
    \end{bmatrix}
\end{equation}

\subsection{Concatenation \& multi-head attention}
The two approaches addressed in the previous sub-sections either directly concatenate a tag vector or derive an attention map to attend a feature map of the ASC system.  
We further propose to apply the two methods together, assuming that the two approaches are complementary (Figure \ref{fig:overall}-(c)).  
Because both methods convert tag vectors before concatenating or deriving an attention map, we can either perform the same transformation for both methods, or apply transformation individually for each method.   
The comparison between the two configurations are described in Table \ref{table:concat_att_fail} and \ref{table:concat_att}.

\section{Experiments and results}
\label{sec:exp}
\subsection{Dataset}
All experiments regarding the ASC task reported in this paper uses the DCASE 2019 task 1-a dataset \cite{DCASE2019Workshop}.  
It comprises of 40 hours of audio recordings from 12 different European cities.  
Ten acoustic scenes are defined and each recording is adjusted to a duration of 10 seconds, making it $14400$ recordings in total.  
Each recording has a sampling rate of $48000$ and recorded in stereo.  
We proceed with the official fold-1 configuration train/test subset split and report the overall classification accuracy on the test set\footnote{Performance using the challenge evaluation set cannot be evaluated because ground truth labels are not released}.  

To train the audio tagging system to extract tag vectors, we use the DCASE 2019 task 2 dataset \cite{Fonseca2019audio}. 
It includes a curated subset of $4970$ audio clips with manual labels. 
The curated subset has a duration range of 0.3 to 30 seconds and a total of 10.5 hours. 
80 sound events are defined. 
The dataset also includes relatively larger noisy subset of $19815$ audio clips with noisy labels. 
The noisy subset has a duration range of 1 to 15 seconds and a total of approximately 80 hours. 

\begin{table}[t]
  \caption{Experimental results of the baseline without the utilization of tag vectors and the three methods using audio tag vectors directly by concatenating to the ASC system. Refer to Figure \ref{fig:overall}-(a) for illustration. \textbf{Bold} depicts the best performance. (-: not applicable)}
  \label{tab:concat}
  \centering
  \begin{tabular}{lccc}
    \toprule
    \multirow{2}{*}{System} &  Transform  &\# Layers     &   \multirow{2}{*}{Acc}\\
    & after concat & before transform \\
    \midrule
    Baseline & - & - & 73.63\\
    \midrule
    Codecat &  $\times$ &  -   &   74.15\\
    \midrule
    \multirow{6}{*}{Before code} & $\circ$ & $\times$   &   74.36\\
    \cmidrule{2-4}
    & $\circ$ &   1   &   74.84\\
    & $\circ$ &   2   &   74.22\\
    & $\circ$ &   \textbf{3}   &   \textbf{75.66}\\
    & $\circ$ &   4   &   75.22\\
    & $\circ$ &   5   &   74.07\\
    \bottomrule
  \end{tabular}
\end{table}

\subsection{Experimental configurations}
The ASC system directly inputs stereo raw waveforms of shape $(479999, 2)$ with pre-emphasis applied. 
Residual blocks include batch normalization \cite{BatchNormalization} and Leaky ReLU activation functions \cite{leaky}. 
The code has a dimensionality of $64$. 
Other details such as convolution layer's filter length are described in Table \ref{tab:dnn_arc}. 

The audio tagging system \cite{Akiyama2019} uses Mel-spectrogram of 128 Mel-frequency channels with augmentations including SpecAugment \cite{park2019specaugment} and slicing.  
Mix-up \cite{zhang2017mixup} is applied to both the ASC system and the audio tagging system.  
ResNet architecture \cite{he2016deep} with approximately 44 million trainable parameters is exploited with few modifications.  
Further details regarding the audio tagging system are addressed in \cite{Akiyama2019}.  

\subsection{Result analysis}
Table \ref{tab:concat} demonstrates the performances of the baseline which does not use tag vectors and the three proposed methods that concatenate a tag vector with an ASC code.  
\textit{Codecat} refers to the first method in Section \ref{sec:proposed}.1 using the concatenation of ASC system's code and tag vector as a new code.  
\textit{Before code} without and with transformation before concatenation refers to the second and the third method respectively.  
Results show that the performance improves in all the proposed configurations.  
The best performance can be achieved by conducting transforms on the tag vector before concatenation and also transforming the concatenated representation to derive the code.  
  
\begin{table}[t]
\caption{Results of conducting multi-head attention using tag vectors. Refer to Figure \ref{fig:overall}-(b) for illustration. \textit{\textbf{Bold}} depicts the best performance from each row.}
\label{table:att}
\centering
\begin{tabular}{c|c|cccc}
\toprule
\multicolumn{2}{c}{} & \multicolumn{4}{c}{\# Transform layers for attention map}\\
\cline{3-6}
\multicolumn{2}{c}{}& 0 & 1 & 2 & 3  \\
\midrule
\multirow{5}{*}{\# Head} & 2 & 75.67 & \textbf{76.58} & 75.93 & 75.50 \\
 & 4 & 75.45 & 75.22 & \textbf{76.24} & 76.17\\
 & 8 & 74.41 & 74.36 & 74.89 & \textbf{75.17}\\
 & 16 & 73.61 & 74.47 & 74.43 & \textbf{75.96}\\
 & 32 & 75.14 & \textbf{75.57}  & 75.24 & 75.31\\
\bottomrule
\end{tabular}
\end{table}

Table \ref{table:att} describes the result of applying multi-head attention to the ASC system with an attention map derived from the tag vector.  
We explore the performance difference according to the different numbers of heads and the number of feature transform layers applied to the tag vectors.  
Results demonstrate that multi-head attention improves the performance more compared to mere concatenation in most configurations.  
The best configurations are achieved by using one layer for tag vector transformation and 2 heads for attention.  

\begin{table}[t]
\caption{Results of applying both concatenation and multi-head attention using tag vectors where tag vector is transformed through separate layers to conduct the two methods. Refer to Figure \ref{fig:overall}-(c) for illustration. \textit{\textbf{Bold}} depicts the best performance from each row.}
\label{table:concat_att_fail}
\centering
\begin{adjustbox}{width=\linewidth}
\makebox[\columnwidth][c]{
\begin{tabular}{c|c|cc|cc}
\toprule
\multicolumn{2}{c}{} & \multicolumn{4}{c}{\# Transform layers for concat/att}\\
\cmidrule(lr){3-6}
\multicolumn{2}{c}{}& 3/3& 3/4 & 4/3 & 4/4\\
\midrule
\multirow{2}{*}{\# Head} & 2 & 74.76 & \textbf{75.74} & 75.17 & 75.14 \\ 
& 4 & 75.31 & 75.84 & \textbf{76.24} & 75.34\\
\bottomrule
\end{tabular}}
\end{adjustbox}
\end{table}

\begin{table}[t]
\caption{Results of applying both concatenation and multi-head attention where tag vector is transformed through separate layers for each method. Refer to Figure \ref{fig:overall}-(c) for illustration. \textit{\textbf{Bold}} depicts the best performance from each row.} 
\label{table:concat_att}
\centering
\begin{tabular}{c|c|cccc}
\toprule
\multicolumn{2}{c}{} & \multicolumn{4}{c}{\# Transform layers}\\
\cline{3-6}
\multicolumn{2}{c}{}& 0 & 1 & 2 & 3  \\
\midrule
\multirow{5}{*}{\# Head} & 2 & \textbf{76.12} & 75.62 & 75.34 & 75.69 \\
 & 4 & 75.43 & 76.00 & 76.24 & \textbf{76.75}\\
 & 8 & 75.84 & 74.71 & \textbf{75.96} & 75.69\\
 & 16 & \textbf{76.00} & 75.81 & 75.91 & 75.72\\
 & 32 & 75.26 & 75.45 & 76.24 & \textbf{76.27}\\
\bottomrule
\end{tabular}
\end{table}

Table \ref{table:concat_att_fail} and \ref{table:concat_att} describes the results of applying both concatenation and multi-head attention.  
Table \ref{table:concat_att_fail} shows the result using a separate fully-connected layers for transforming a tag vector for concatenation and attention map, whereas, Table \ref{table:concat_att} shows the result using shared layers for the two methods.  
For experiments in Table \ref{table:concat_att_fail}, the number of heads of 2 and 4 was explored for multi-head attention and the number of transform layers of 3 and 4 was explored because these configurations demonstrated best performances in Table \ref{tab:concat} and \ref{table:att}.  

Results show that using separate transform layers (76.24 \%) leads to worse performance than using only multi-head attention (76.57 \%).  
Using identical layers for tag vector transformation and conducting both concatenation and attention resulted in further performance improvement, showing an accuracy of 76.75 \%.  
In our analysis, separate transformation layers led to worse performance because of overfitting incurred by too many parameters of fully-connected layers.  

\begin{table}[t]
\caption{Comparison of the proposed system with top three performing system using raw waveform as the input on the DCASE 2019 challenge. (-: not reported)}
\label{tab:sota}
\centering
\setlength{\tabcolsep}{11pt}
\begin{tabular}{lcc}
\toprule
System & \# Param & Acc (\%)\\
\midrule
Huang \textit{et. al.} \cite{Huang2019} & 53452k & 76.08\\
Jung \textit{et. al.} \cite{jung2019distilling} & 636k & 75.81\\
Zheng \textit{et. al.} \cite{Zheng2019} & - & 69.23\\
\textbf{Ours} & 676k & \textbf{76.75}\\
\bottomrule
\end{tabular}
\end{table}

Finally, Table \ref{tab:sota} shows the comparison between the best performing system of our study and three systems that demonstrated the best performance using raw waveform as input.  
Huang \textit{et. al.}'s system utilized SincNet architecture \cite{ravanelli2018speaker} which replaces the first convolution layer with the proposed sinc-convolution layers combined with AclNet \cite{huang2018aclnet}.  
Jung \textit{et. al.}'s system utilized teacher-student learning \cite{li2014learning} with specialist DNNs \cite{hinton2015distilling}.  
Zheng \textit{et. al.}'s system utilized an end-to-end DNN architecture with random cropping and padding augmentations.  
Results show that the proposed system shows the best performance among other systems that use raw waveform as input.  

\section{Conclusion} 
In this paper, we explored various approaches to exploit audio tagging representations for improving the ASC system which mimics human perception of the ASC task.  
Three methods were studied to utilize audio tagging representation through a tag vector concatenation.  
Multi-head attention was adopted to emphasize the feature map of the ASC system using a tag vector.  
Based on the success of both concatenation and multi-head attention, we achieved further performance improvement by applying the proposed methods concurrently.  
This system showed an accuracy of 76.75 \% on the fold-1 test set configuration of the DCASE task 1-a dataset.

\section{Acknowledgements}
This research was supported by Basic Science Research Program through the National Research Foundation of Korea(NRF) funded by the Ministry of Science, ICT \& Future Planning(2020R1A2C1007081).

\newpage
\bibliographystyle{IEEEtran}
\bibliography{refs}
\end{document}